\documentclass[aps,twocolumn,pre,showpacs,floatfix,superscriptaddress]{revtex4}

\usepackage{amsmath}
\usepackage{graphicx}
\usepackage{amssymb}
\usepackage{bm}
\usepackage{longtable}
\usepackage{color}
\definecolor{darkgreen}{rgb}{0,0.5,0}
\begin{document}

\title{ Microcanonical finite-size scaling in specific heat
  diverging 2$^\text{nd}$ order phase transitions}
\author{L.A.~Fernandez}
\affiliation{Departamento de F\'\i{}sica Te\'orica I, 
  Universidad Complutense, 28040 Madrid, Spain.}
\affiliation{Instituto de Biocomputaci\'on and
  F\'{\i}sica de Sistemas Complejos (BIFI), 50009 Zaragoza, Spain.}

\author{A. Gordillo-Guerrero} 
\affiliation{Departamento de Ingenier\'{\i}a El\'ectrica, 
  Electr\'onica y Autom\'atica,
  Universidad de Extremadura, 10071 Caceres, Spain.}
\affiliation{Instituto de Biocomputaci\'on and
  F\'{\i}sica de Sistemas Complejos (BIFI), 50009 Zaragoza, Spain.}

\author{V. Martin-Mayor}
\affiliation{Departamento de F\'\i{}sica Te\'orica I, 
  Universidad Complutense, 28040 Madrid, Spain.}
\affiliation{Instituto de Biocomputaci\'on and
  F\'{\i}sica de Sistemas Complejos (BIFI), 50009 Zaragoza, Spain.}

\author{J.J. Ruiz-Lorenzo} 
\affiliation{Departamento de F\'{\i}sica,
  Universidad de Extremadura, 06071 Badajoz, Spain.}
\affiliation{Instituto de Biocomputaci\'on and
  F\'{\i}sica de Sistemas Complejos (BIFI), 50009 Zaragoza, Spain.}

\date{\today}

\begin{abstract}
A Microcanonical Finite Site Ansatz in terms of quantities measurable
in a Finite Lattice allows to extend phenomenological renormalization
(the so called quotients method) to the microcanonical
ensemble. The Ansatz is tested numerically in two models where the
canonical specific-heat diverges at criticality, thus implying
Fisher-renormalization of the critical exponents: the $3D$
ferromagnetic Ising model and the $2D$ four-states Potts model (where
large logarithmic corrections are known to occur in the canonical
ensemble). A recently proposed microcanonical cluster method allows to
simulate systems as large as $L=1024$ (Potts) or $L=128$ (Ising). The
quotients method provides extremely accurate determinations of the
anomalous dimension and of the (Fisher-renormalized) thermal $\nu$
exponent. While in the Ising model the numerical agreement with our
theoretical expectations is impressive, in the Potts case we need to
carefully incorporate logarithmic corrections to the microcanonical
Ansatz in order to rationalize our data.
\end{abstract}
\maketitle 

\section{Introduction}
The canonical ensemble enjoys a predominant position in Theoretical
Physics due to its many technical advantages (convex effective
potential on finite systems, easily derived Fluctuation-Dissipation
theorems, etc.)\footnote{Depending on context, sometimes the
  grand-canonical and canonical ensembles are on the same relative
  position than the canonical and microcanonical ones.}. This somehow
arbitrary choice of ensemble is justified by the Ensemble Equivalence
property, that holds in the Thermodynamic Limit for systems with short
range interactions.

However, in spite of this long standing prejudice in favor of the
canonical ensemble, the canonical analysis of phase transitions is
{\em not} simpler. The advantages of microcanonical analysis of
first-order phase transitions has long been
known~\cite{FIRST-ORDER,VICTORMICRO}, and indeed become overwhelming
in the study of disordered systems~\cite{POTTS3D}.  Furthermore, the
current interest in mesoscopic or even nanoscopic systems, where
Ensemble Equivalence does not hold, provides ample motivation to study
other statistical ensembles and, in particular, the microcanonical
one~\cite{GROSS}. Besides, microcanonical Monte Carlo~\cite{LUSTIG} is
now as simple and efficient as its canonical counterpart (even
microcanonical cluster algorithms are known~\cite{VICTORMICRO}). Under
such circumstances, it is of major interest the extension to the
microcanonical framework of Finite-Size Scaling
(FSS)~\cite{BINDER,BARBER,PRIVMAN,VICTORAMIT} for systems undergoing a
continuous phase transition.

The relation between the microcanonical and the canonical critical
behavior is well understood only in the Thermodynamic Limit. A global
constraint modifies the critical exponents, but only if the specific-heat
of the unconstrained system diverges with a
positive critical exponent $\alpha>0$~\cite{FISHER-RENORM} (however, see~\cite{DOHM}).
The modification in the
critical exponents, named Fisher renormalization, is very simple.  Let
$L$ be the system size, and consider an observable $O$ (for instance,
the correlation length) whose scaling behavior in the infinite-volume
canonical system is
\begin{equation}
\langle O \rangle_{L=\infty,T}^\text{canonical} \propto |t|^{-x_O}\,,\quad
t=\frac{T-T_\text{c}}{T_\text{c}}\,.\label{EXPO-CANONICO}
\end{equation}
Now, let $e$ be the internal energy density and $e_\text{c}=\langle
e\rangle^\text{canonical}_{L=\infty,T_\text{c}}$. Consider the
microcanonical expectation value of the {\em same} observable $O$ in
\eqref{EXPO-CANONICO}, but now at fixed energy $e$. The scaling behavior
\eqref{EXPO-CANONICO} translates to
\footnote{In the particular case of the fixed-energy constraint,
Eq.~\eqref{EXPO-MICRO-CANONICO} follows from \eqref{EXPO-CANONICO} and from
the Ensemble Equivalence property
$$\langle O\rangle_{L=\infty,e}=\langle
O\rangle_{L=\infty,T}^\text{canonical}\,,\quad\text{if }\quad e=\langle
e\rangle_{L=\infty,T}^\text{canonical}\,.$$ Indeed, it suffices to
notice that ($C(T)$ is the canonical specific heat, $C\propto
|t|^{-\alpha}$),
$$e-e_\text{c}=\int_{T_\text{c}}^T\,\text{d}T\, C(T)\propto |t|^{1-\alpha}\text{ or } |t|\propto |e-e_\text{c}|^{\frac{1}{1-\alpha}}\,.$$}
\begin{equation}
\langle O\rangle_{L=\infty, e} \propto |e-e_\text{c}|^{-x_{O,\text{m}}}\,,\quad
x_{O,\text{m}}=\frac{x_O}{1-\alpha}\,.
\label{EXPO-MICRO-CANONICO}
\end{equation}
We will denote the microcanonical exponents with the subindex ``m''.
Hence, the Fisher renormalization of the correlation length exponent
$\nu$, is $\nu\rightarrow \nu_\text{m}=\nu/(1-\alpha)$, that of the
order parameter exponent is $\beta\rightarrow
\beta_\text{m}=\beta/(1-\alpha)$, etc. On the other hand, the
anomalous dimension $\eta=\eta_\text{m}$ is invariant under
Fisher renormalization~\cite{FISHER-RENORM}. See also~\cite{KENNA},
for a recent extension of Fisher renormalization
to the case of {\em logarithmic} scaling corrections.

As for systems of finite size, the microcanonical
FSS~\cite{MFSS1,MFSS2,MFSS3} is at the level of an Ansatz. This Ansatz
is obtained from the canonical one merely by replacing the free-energy
density by the entropy density, and using Fisher renormalized critical
exponents.  The microcanonical Ansatz reproduces the canonical
one~\cite{KASTNER2001}, and it has been subject of some numerical
testing~\cite{MFSS3,BEHRINGER2006}. Furthermore, systems undergoing
Fisher Renormalization (due to a global constraint other than the
energy) do seem to obey FSS as well~\cite{TROSTER08}.

A difficulty lies in the fact that the current forms of the microcanonical FSS
Ansatz~\cite{MFSS1,MFSS2,MFSS3} are in a somehow old-fashioned form.
Indeed, they are formulated in terms of quantities such as
$e_\text{c}$ or the critical exponents, which are not accessible in
the absence of an analytical solution. In this respect, a great step
forward was achieved in a canonical context~\cite{FSS-PISA} when it
was realized that the Finite-Lattice correlation length~\cite{COOPER}
allows to formulate the FSS Ansatz in terms of quantities computable
in a Finite-Lattice. This formulation made practical to extend
Nightingale's phenomenological renormalization~\cite{NIGHTINGALE} to
space dimensions $D>2$ (the so-called quotients
method~\cite{QUOTIENTS}).

Here, we will extend the microcanonical FSS Ansatz to a modern form,
allowing us to use the quotients method. We will test numerically this
extended Ansatz in two models with $\alpha>0$, hence undergoing
nontrivial Fisher Renormalization, namely the $D=3$ ferromagnetic
Ising model, and the $D=2$ four-states ferromagnetic Potts model. The Potts
model has the added interest of suffering, in its canonical form,
quite strong {\em logarithmic} corrections to scaling that are
nevertheless under relatively strong analytical
control~\cite{Salas}. It will be, therefore, quite a challenge
to control the logarithmic corrections in the microcanonical setting.

The layout of the rest of this paper is as follows.  In
Sect.~\ref{MICRO-SECT}, we briefly recall the particular
microcanonical ensemble used in this work (Lustig's microcanonical set
up~\cite{LUSTIG}, where the Fluctuation-Dissipation formalism
of~\cite{VICTORMICRO} applies).  In Sect.~\ref{MFSSA-SECT} we present
our extended microcanonical FSS Ansatz.  A brief description of
simulated models and measured observable is presented in
Sect.~\ref{MODELS_OBS-SECT} while the specific simulation details are
given in Sect.~\ref{SIMU-SECT}. The results both for the $D=3$ Ising
model and for the $D=2$ Potts model are given in
Sect.~\ref{ISING3D-SECT} and~\ref{POTTS2D-SECT} respectively. Finally
we devote Sect.~\ref{CONC-SECT} to the conclusions.  In addition, 
in Appendix~\ref{AppendixA} we propose an
extension of the quotients method, aimed to speed up convergence to the large
$L$ limit in the presence of multiplicative logarithmic corrections.

\section{The microcanonical ensemble}\label{MICRO-SECT}

The first step in the construction of the ensemble is an extension of
the configuration space.  We add $N(=L^D)$ real momenta, $p_i$, to our
$N$ original variables, $\sigma_i$ (named spins
here)~\cite{LUSTIG,VICTORMICRO}.  Note that this extended
configuration, $\{\sigma_i,p_i\}$, appears in many numerical schemes
(consider, for instance, Hybrid Monte Carlo~\cite{HYBRID} simulations
in Lattice Gauge Theory). We shall  work in
the {\em microcanonical} ensemble for the $\{\sigma_i,p_i\}$ system.

Let ${\cal U}$ be the original spin Hamiltonian
(e.g. Eq.~\eqref{SPIN-HAMILTONIAN} in our case).  Our total energy
is~\footnote{Note that this microcanonical ensemble exactly matches
  the conditions in the original Fisher work~\cite{FISHER-RENORM}: the
  momenta are some {\em hidden} degrees of freedom in thermal
  equilibrium with the spins, and a global constraint is imposed. It
  is also amusing to rederive the results in Sect.~\ref{MICRO-SECT}
  considering $\varGamma$ momenta per spin (in this work $\varGamma=1$, while Lustig~\cite{LUSTIG} always considered $\varGamma=3$). If
  one takes the limit $\varGamma\to\infty$, at fixed $N,$ the canonical
  probability is recovered for the spins.}
\begin{equation}
{\cal E}= \sum_{i=1}^{N} \frac{p_i^2}{2}\ +\ {\cal U}\, \quad (e\equiv {\cal
E}/N\,,\ u\equiv {\cal U}/N)\,.\label{ENERGIATOTAL}
\end{equation}
The momenta contribution,
\begin{equation}
N\kappa \equiv \sum_{i=1}^{N} \frac{p_i^2}{2}\,,
\end{equation}
is necessarily positive, and it is best thought of as a ``kinetic''
energy.  In this mechanical analog, the original spin Hamiltonian
${\cal U}$ can be regarded as a ``potential'' energy.

The canonical partition function is ($\beta\!\equiv\!1/T$)
\begin{equation}
Z_N(\beta)=\int_{-\infty}^\infty\prod_{i=1}^N\text{d}p_i\sum_{\{\sigma_i\}}\
\text{e}^{-\beta{\cal E}}=\left(\frac{2\pi}{\beta}\right)^{\frac{N}{2}} 
\sum_{\{\sigma_i\}}\ \text{e}^{-\beta{\cal U}}\,,
\end{equation}
where $\sum_{\{\sigma_i\}}$ denotes summation over
spin configurations. Hence, the $\{p_i\}$ play the role of a Gaussian thermostat. The
$\{p_i\}$ are statistically uncorrelated with the spins. Since
$\langle \kappa \rangle^\text{canonical}_{L,\beta}=1/(2\beta)$, one has
$\langle e\rangle^\text{canonical}_\beta= \langle
u\rangle^\text{canonical}_\beta + 1/(2\beta)\,$.

Furthermore, given the statistical independence of $\kappa$ and $u$,
the canonical probability distribution function for $e$,
$P_\beta^{(L)}(e)$, is merely the convolution of the distributions for
$\kappa$ and $u$:
\begin{equation}
P_\beta^{(L)}(e)=\int_0^\infty\,\text{d}\,\kappa\  P_\beta^{(L),\kappa}(\kappa)\,
P_\beta^{(L),u}(e-\kappa)\,.
\end{equation}
In particular, note that for spin systems on a finite lattice,
$P_\beta^{(L),u}(u)$ is a sum of (order $N$) Dirac's $\delta$
functions.  Now, since the canonical variance of $\kappa$ is $1/(\beta
\sqrt{2 N})$, roughly $\sqrt{N}$ discrete $u$-levels, with $u\sim
e-1/(2\beta)$, give the most significant contribution to
$P_\beta^{(L)}(e)$. We see that the momenta's kinetic energy  
provide a natural smoothing of the comb-like
$P_\beta^{(L),u}(u)$. Once we have a conveniently smoothed
$P_\beta^{(L)}(e)$, we may proceed to the definition of the entropy.

In a microcanonical setting, the crucial role is played by the entropy
density, $s(e,N)$, given by 
\begin{equation}
\exp[N s(e,N)]=
\int_{-\infty}^\infty\prod_{i=1}^N\text{d}p_i\sum_{\{\sigma_i\}}\ 
\delta(Ne-\cal{E})\,.\label{MICRO1}
\end{equation}

Integrating out the $\{p_i\}$  using Dirac's delta function in (\ref{MICRO1})
we get
\begin{eqnarray}
\exp[N s(e,N)] &=&\displaystyle \frac{(2\pi N)^{\frac{N}{2}}}{N \Gamma(N/2)} \sum_{\{\sigma_i\}}
\omega(e,u,N)\,,\label{MICRO2}\\
\omega(e,u,N)&\equiv&(e-u)^{\frac{N-2}{2}}\theta(e-u)\;.\label{MICROPESO}
\end{eqnarray}
The step
function, $\theta(e-u)$, enforces $e>u$.  Eq.~\eqref{MICRO2} suggests to
define the microcanonical average at fixed $e$ of any function of $e$
and the spins, $O(e,\{\sigma_i\})$, as~\cite{LUSTIG}
\begin{equation}
\langle O\rangle_e\equiv
\frac{\sum_{\{\sigma_i\}}\,O(e,\{\sigma_i\})\,\omega(e,u,N)}{{\sum_{\{\sigma_i\}}\omega(e,u,N)}}\,.\label{MICROPROB}
\end{equation}

We use Eq.~\eqref{MICRO2} to compute $\text{d}s/\text{d}e$:~\cite{VICTORMICRO} 
\begin{eqnarray}
\frac{\text{d} s(e,N)}{\text{d} e}&=
&\langle\hat\beta(e;\{\sigma_i\})\rangle_e\,,\label{FD1}\\
\hat\beta(e;\{\sigma_i\})&\equiv&\frac{N-2}{2N (e -u)}\,.
\end{eqnarray}

Keeping in mind the crucial role of the generating-functional in
Field-Theory (see e.g.~\cite{VICTORAMIT}), we extend the definition
\eqref{MICRO1} by considering a linear coupling between the spins and
a site dependent source field $h_i$:
\begin{equation}
\exp[N s(e,\{h_i\},N)]=
\int_{-\infty}^\infty\prod_{i=1}^N\text{d}p_i\sum_{\{\sigma_i\}}\ \text{e}^{\sum_i\, h_i \sigma_i}
\delta(Ne-\cal{E})\,,\label{MICRO-FUENTE}
\end{equation}
where ${\cal E}=Ne$ is still given by Eq.~\eqref{ENERGIATOTAL},
without including the source term.
In this way, the microcanonical spin correlation functions follow from derivatives of $s(e,\{h_i\},N)$:
\begin{eqnarray}
\left.\frac{\partial [N\, s]}{\partial h_k}\right|_{e,\{h_i\},N}&=&\langle \sigma_k \rangle_{e, \{h_i\}}\,,\\\nonumber
\left.\frac{\partial^2 [N\, s]}{\partial h_k\partial h_l}\right|_{e,\{h_i\},N}&=&\langle \sigma_k \sigma_l \rangle_{e, \{h_i\}}- 
\langle \sigma_k\rangle_{e, \{h_i\}}\, \langle\sigma_l \rangle_{e, \{h_i\}}\,.
\end{eqnarray}
In particular, if the source term is uniform $h_i=h$ we observe that
the microcanonical susceptibility is given by standard
fluctuation-dissipation relations, see Ref.~\onlinecite{VICTORAMIT} and
Eq.~(\ref{chi-micro-def}), below.

\subsection{Ensemble equivalence}

Eq.~\eqref{MICRO1} ensures that the {\em
  canonical} probability density function for $e$ is
\begin{equation}
P_{\beta}^{(L)}(e)=\frac{N}{Z_N(\beta)}\exp[N(s(e,N)-\beta e)]\,,\label{LINK0}
\end{equation}
hence, Eq.~\eqref{FD1},
\begin{equation}
\log P_{\beta}^{(L)}(e_2)-\log P_{\beta}^{(L)}(e_1)=
N\int_{e_1}^{e_2}\text{d}e\, \left(
\langle\hat\beta\rangle_e -\beta\right)\,.\label{LINK1}
\end{equation}

The relation between the canonical and the microcanonical spin-values is given
by 
\begin{equation}
\langle O \rangle^\text{canonical}_\beta=\int_{-\infty}^\infty\,\text{d}e\ 
\langle O\rangle_e\, P_{\beta}^{(L)}(e)\,.\label{LINK2}
\end{equation}
Now, Eqs.~\eqref{LINK0} and~\eqref{LINK2} imply that the canonical mean-value will
be dominated by a saddle-point at $e^\text{SP}$,
\begin{equation}
\langle \hat\beta \rangle_{e^\text{SP}_{L,\beta}}=\beta\,,\label{LINK3}
\end{equation}
which can be read as yet another expression of Thermodynamics second-law,
$T\text{d}s=\text{d}e$\,.

The condition of thermodynamic stability (namely that $\langle
\hat\beta \rangle_e $ be a monotonically decreasing function of $e$)
ensures that the saddle point is unique and that $e^\text{SP}$ is a maximum of
$P_\beta(e)$. Under the thermodynamic stability condition and if, in
the large $L$ limit, 
\begin{equation}
\left.\frac{{\text d} \langle \hat\beta\rangle_e}{{\text d} e}\right|_{e^\text{SP}_{L,\beta}} <0\,,\label{LINK4}
\end{equation}
the saddle point approximation becomes exact:
\begin{equation}
e^\text{SP}_{L=\infty,\beta} = \langle e\rangle_{L=\infty,\beta}^\text{canonical}\,,\label{LINK5}
\end{equation}
and we have Ensemble Equivalence:
\begin{equation}
\langle O\rangle_{L=\infty,e^\text{SP}_{L=\infty,\beta}}=\langle O\rangle^\text{canonical}_{L=\infty,\beta}\,.\label{LINK6}
\end{equation}
It follows that $1/[{\text d} \langle \hat\beta \rangle_e/{\text
    d} e]$ at $e^\text{SP}_{L=\infty,\beta}$ will tend in the
large-$L$ limit to minus the canonical specific heat. Thus, if
the critical exponent $\alpha$ is positive, Eq.~\eqref{LINK4} will fail
precisely at $e_\text{c}$. Hence, Eq.~\eqref{LINK6} can be expected to hold
for all $e$ but $e_\text{c}$ (or for all $\beta$ but $\beta_\text{c}$).

\subsection{Double peaked histogram}{\protect\label{SECT-DOS-PICOS}}
The situation can be slightly more complicated if
$P_{\beta_\text{c}}(e)$ presented two local maxima, remindful of phase
coexistence. This is actually the case for one of our models, the
$D\!=\!2$, four states Potts model~\cite{Fukugita}. From
Eq.~\eqref{LINK1} it is clear that the solution to the saddle point
equation~\eqref{LINK3} will no longer be unique. We borrow the
following definitions from the analysis of first-order phase
transitions (where true phase coexistence takes
place)~\cite{VICTORMICRO}:
\begin{itemize}
\item The rightmost root of~\eqref{LINK3}, $e_{L,\beta}^\text{d}$, is a local
maximum of $P_{\beta}^{(L)}$ corresponding to the ``disordered phase''.
\item The leftmost root of~\eqref{LINK3}, $e_{L,\beta}^\text{o}$, is a local
maximum of $P_{\beta}^{(L)}$ corresponding to the ``ordered phase''.
\item The second rightmost root of~\eqref{LINK3}, $e_{L,\beta}^*$  is a local
minimum of $P_{\beta}^{(L)}$.
\end{itemize}
Maxwell's construction yields the finite-system  critical
point, $\beta_{\text{c},L}$ (see Fig.~\ref{beta_e_L}):
\begin{equation}
0=\int_{e^\text{o}_{L,{\beta_{\text{c},L}}}}^{e^\text{d}_{L,{\beta_{\text{c},L}}}}
\text{d}e\, \left(\langle\hat\beta\rangle_e -\beta_{\text{c},L}\right)\,,\label{MAXWELL}
\end{equation}
and the finite-system estimator of the ``surface tension''
\begin{equation}
\Sigma^L=\frac{N}{2L^{D-1}} \int_{e^*_{L,{\beta_{\text{c},L}}}}^{e^\text{d}_{L,{\beta_{\text{c},L}}}}
\text{d}e\, \left(
    \langle\hat\beta\rangle_e -\beta_{\text{c},L}\right)\,.\label{SIGMAEQ}
\end{equation}
Of course, in the large-$L$ limit and for a continuous transition,
$\Sigma^L\to 0$, $\beta_\text{c}^L\to \beta_\text{c}$ and
$e^\text{d}_{L,{\beta_{\text{c},L}}},e^\text{o}_{L,{\beta_{\text{c},L}}}\to
e_\text{c}\,$.

\section{Our Microcanonical Finite-Size Scaling Ansatz}
\label{MFSSA-SECT}
Usually, the Microcanonical FSS Ansatz takes the form of a scaling
form for the entropy density~\cite{MFSS1,MFSS2,MFSS3}. In close
analogy with the canonical case, one assumes that $s(e,\{h_{\vec
  x}\},N)$ can be divided in a regular part, and a singular term
$s_\text{sing}(e,\{h_{\vec x}\},N)$. The regular part is supposed to
converge for large $L$ (recall that $N=L^D$) to a smooth function of
its arguments. Hence, all critical behavior comes from
$s_\text{sing}(e,\{h_{\vec x}\},N)$. Note as well that we write
$\{h_{\vec x}\}$, instead of $\{h_i\}$, to emphasize the spatial
dependence of the sources (supposedly very mild~\cite{VICTORAMIT}).
Hence,
\begin{equation}
s_{\text{sing}}(e,\{h_{\vec x}\},N) = L^{-D} g \left(L^{\frac{1}{\nu_\text{m}}}(e-e_\text{c}), \{ L^{y_h}h_{\vec x}\} \right)\,.\label{FSSA-OLD}
\end{equation}
Here, $g$ is a very smooth function of its arguments, while
$y_h=1+\frac{D-\eta}{2}$ is the canonical exponent, see
e.g.~\cite{VICTORAMIT}, which does not get
Fisher-renormalized. Scaling corrections due to irrelevant scaling
fields, have been ignored by other authors~\cite{MFSS1,MFSS2,MFSS3},
but will be important for our precision tests. We will propose here
alternative forms of the Ansatz~\eqref{FSSA-OLD}, more suitable
for a numerical work where neither $e_\text{c}$ nor the critical
exponents are known beforehand.

Our first building block is the infinite-system microcanonical
correlation length, $\xi_{\infty,e}$.  Indeed, Ensemble Equivalence implies
that, in an infinite system, the long-distance behavior of the
microcanonical spin-spin propagator $G(\vec r;e)=\langle\sigma_{\vec
  x} \sigma_{\vec x+\vec r}\rangle_e - \langle\sigma_{\vec x}\rangle_e
\langle \sigma_{\vec x+\vec r}\rangle_e $ behaves for large $\vec r$
as in the canonical ensemble (close to a critical point
$\xi_{\infty,e}$ is large, so that rotational invariance is recovered
in our lattice systems):
\begin{equation}
G(\vec r;e) =\frac{A}{r^{D-2+\eta}} \text{e}^{-r/\xi_{\infty,e}}\,,
\end{equation} 
where A is a constant.
In particular, note that Ensemble-Equivalence implies that the
anomalous dimension $\eta$ does not get Fisher-renormalized. We expect
$\xi_{\infty,e}=\xi^\text{canonical}_{\infty,T}$ if the correspondence
between $e$ and $T$ are fixed through
$e=\langle e \rangle^\text{canonical}_{L=\infty,T}\,.$

The basic assumption underlying the FSS Ansatz is that the approach to the
$L\to\infty$ limit is governed by the dimensionless ratio
$L/\xi_{\infty,e}$. Hence, our first form of the Ansatz for the observable $O$
whose critical behavior was discussed in Eq.~\eqref{EXPO-MICRO-CANONICO}
is
\begin{equation}
\langle O\rangle_{L,e}= L^{\frac{x_{O,\text{m}}}{\nu_\text{m}}} f_O(L/\xi_{\infty,e})+\ldots\,.\label{FSSA1}
\end{equation}
In the above, the dots stand for scaling-corrections, while the
function $f_O$ is expected to be very smooth (i.e. differentiable to a
large degree or even analytical). A second form of the Ansatz is
obtained by substituting the scaling behavior $\xi_{\infty,e}\propto
|e-e_\text{c}|^{-\nu_\text{m}}$:
\begin{equation}
\langle O\rangle_{L,e}= L^{\frac{x_{O,\text{m}}}{\nu_\text{m}}} \tilde f_O\left (L^{1/\nu_\text{m}} (e-e_\text{c})\right)+\ldots\,.\label{FSSA2}
\end{equation}
Again, $\tilde f_O$ is expected to be an extremely smooth function of
its argument~\footnote{Note that the microcanonical weight (\ref{MICROPESO}) is {\em not} analytical at each energy level of the spin Hamiltonian.}. In particular, this is the form of the Ansatz that
follows from Eq.~\eqref{FSSA-OLD} by derivating with respect to $e$ or
from the source terms.

However, the most useful form of the Ansatz is obtained by applying
\eqref{FSSA1} to the Finite-Lattice correlation length $\xi_{L,e}$,
obtained in a standard way (see Ref.~\cite{VICTORAMIT}) from the
finite-lattice microcanonical propagator.  We expect $\xi_{L,e}/L$ to
be a smooth, one-to-one function of $L/\xi_{\infty,e}$, that can be
inverted to yield $L/\xi_{\infty,e}$ as a function of
$\xi_{L,e}/L$. Hence, our preferred form of the FSS Ansatz is
\begin{equation}
\langle O\rangle_{L,e}= L^{\frac{x_{O,\text{m}}}{\nu_\text{m}}}\left[F_O\left(\frac{\xi_{L,e}}{L}\right)+L^{-\omega} G_O\left(\frac{\xi_{L,e}}{L}\right)+\ldots\right]\label{FSSA3}\,.
\end{equation}
Here, $F_O$ and $G_O$ are smooth functions of their arguments and
$\omega$ is the first Universal Scaling Corrections exponent.

It is important to note that exponent $\omega$ does not get
Fisher-renormalized. Indeed,  let us
consider an observable $O$ with critical exponent $x_O$ at a
temperature $T$ such $e=\langle
e\rangle_{L=\infty,T}^\text{canonical}$. Now, ensemble equivalence
tells us that $O_{L=\infty,T}^\text{canonical}=O_{L=\infty,e}$ and
that $\xi_{L=\infty,T}^\text{canonical}=\xi_{L=\infty,e}$. Eliminating
$T$ in favor of $\xi_{L=\infty,T}^\text{canonical}$, see e.g.~\cite{VICTORAMIT},
we have
\begin{equation}
O_{L=\infty,T}^\text{canonical} = \xi_{L=\infty,e}^{x_O/\nu}[A_0 + B_O\xi_{L=\infty,e}^{-\omega}+\ldots]\,, 
\end{equation}
where $A_0$ and $B_0$ are scaling amplitudes. It follows that
$\omega_\text{m}=\omega$, and that
$x_O/\nu=x_{O,\text{m}}/\nu_\text{m}$.

\subsection{The quotients method}\label{subsec:quotients}

Once we have Eq.~\eqref{FSSA3} in our hands, it is straightforward to
generalize the quotients method~\cite{QUOTIENTS}. In
Appendix~\ref{AppendixA} we describe how it should be modified in the
presence of (multiplicative) logarithmic corrections to scaling.

Let us compare data obtained {\em at the same} value of $e$ for a pair of
lattices $L_1=L$ and $L_2=sL$ with $s>1$. We expect that a single
$e_{\text{c},L_1,L_2}$ exists such that the correlation-length in units of the
lattice size coincides for both systems:
\begin{equation}
\frac{\xi_{L,e_{\text{c},L_1,L_2}}}{L}=\frac{\xi_{sL,e_{\text{c},L_1,L_2}}}{sL}\,.\label{QUO1}
\end{equation}
Hence, if we compare now in the two lattices the observable $O$ in \eqref{FSSA3}, precisely at $e_{\text{c},L,sL}$, we have
\begin{equation}
\frac{\langle O\rangle_{sL,e_{\text{c},L_1,L_2}}}{\langle O\rangle_{L,e_{\text{c},L_1,L_2}}}=
s^{\frac{x_{O,\text{m}}}{\nu_\text{m}}}\left[1 + A_{O,s}L^{-\omega} +\ldots\right],\label{QUO2}
\end{equation}
where $A_{O,s}$ is a non-universal scaling amplitude. One
considers this equation for fixed $s$ (typically $s=2$), and uses it to
extrapolate to $L=\infty$ the $L$-dependent estimate of the critical
exponents ratio $x_{O,\text{m}}/\nu_\text{m}\,$. At the purely
numerical level, mind as well that there are strong statistical
correlations between the quotients in \eqref{QUO1} and in
\eqref{QUO2}, that can be used via a jackknife method (see
e.g.~\cite{VICTORAMIT}) to strongly reduce the statistical errors in
the estimate of critical exponents.

In this work, we shall compute the critical exponents from the following operators
($\chi$ is the susceptibility, while $\xi$ is the correlation length, see Sect.~\ref{MODELS_OBS-SECT}
for definitions):
\begin{eqnarray}
\chi & \rightarrow & x_O= \nu_\text{m} (2-\eta)\,,\\
\partial_e \xi & \rightarrow & x_O= \nu_\text{m} + 1\,.
\end{eqnarray}

As for the $L$ dependence of $e_{\text{c},L,s}$, it follows from Eq.~\eqref{FSSA2} as applied to $\xi_L/L$ for the two lattice sizes $L$ and $sL$~\cite{BINDER,VICTORAMIT}:
\begin{equation}
e_{\text{c},L,s}=e_\text{c}+ B \frac{1-s^{-\omega}}{s^{1/\nu_\text{m}}-1}L^{-(\omega+\frac{1}{\nu_\text{m}})}+\ldots\label{SHIFT-ec-LEADING}
\end{equation}
($B$ is again a non-universal scaling amplitude). In particular, if
one works at fixed $s$, $e_{\text{c},L,sL}$ tends to $e_\text{c}$ for
large $L$ as $L^{-(\omega+\frac{1}{\nu_\text{m}})}$~\footnote{Note
  that, Eq.~\eqref{FSSA2} tells us that, if the energy histogram is
  double-peaked, see Sect.~\ref{SECT-DOS-PICOS}, the histogram maxima
  will tend to $e_\text{c}$ only as $L^{-1/\nu_\text{m}}\,$.}.

\section{Models and Observables}
\label{MODELS_OBS-SECT}
We will define here the Model and Observables of a generic $D$-dimensional
$Q$-states Potts model. The numerical study has been done for two instances of
this model: the three dimensional Ising ($Q\!=\!2$) model, and the two
dimensional $Q\!=\!4$ Potts model.

We place the spins $\sigma_i=1,\ldots,Q$ at the nodes of a hypercubic
$D$-dimensional lattice with linear size $L$ and periodic boundary
conditions.

The Hamiltonian is
\begin{equation}
{\cal U}=-\sum_{<i,j>}
\delta_{\sigma_i\sigma_j}\ ,\label{SPIN-HAMILTONIAN}
\end{equation}
where $<i,j>$ denotes first nearest neighbors. 
For a given spin, $\sigma$, we define the normalized $Q$-vector $\vec s$, 
whose $q$-th component is
\begin{equation}
s_q=\sqrt{\frac{Q}{Q-1}}\Big(\delta_{\sigma q}-\frac1Q\Big)\;.
\end{equation}
A $Q$ components order parameter for the ferromagnetic transition is
\begin{equation}
\vec{\cal M}=\frac{1}{L^D}\sum_{i}\vec s_i\;, 
\end{equation}
where $i$ runs over all the lattice sites. We will now consider microcanonical averages.
The spatial correlation function is
\begin{equation}
\begin{array}{rcl}
C({\bm r}'-{\bm r})&=&
\Big\langle \vec s({\bm r})\cdot\vec s({\bm r}')\Big\rangle_{\!e}\\
&=&\displaystyle\frac{Q}{Q-1}\Big\langle \delta_{\sigma({\bm
    r})\sigma({\bm r}')} - \frac1Q\Big\rangle_e \;.
\end{array}
\end{equation}
Our definition for the correlation length at a given internal energy density
$e$, is computed from the Fourier transform of $C$ 
\begin{equation}
\hat C({\bm k})=
\sum_{\bm r} C({\bm r})\,{\text e}^{{\text i}{\bm k}\cdot{\bm r}}\;,\label{hatC-def}
\end{equation}
at zero and minimal ($\Vert {\bm k}_\text{min}\Vert=2\pi/L$) momentum~\cite{COOPER,VICTORAMIT}:
\begin{equation}
\xi(e,L)=\frac{\sqrt{\hat C(0)/\hat C({\bm
      k}_{\text{min}})-1}}{2\sin(\pi/L)} \;.
\label{xi-def}
\end{equation}
Note that  $\hat C$ can be easily computed
in terms of the Fourier transform of the spin field, $\hat s({\bm k})$, as
\begin{equation}
\hat C({\bm k})=L^D \big\langle \hat s({\bm k})\cdot\hat s({-\bm k})\big\rangle_e\;,
\end{equation}
and that the microcanonical magnetic susceptibility is
\begin{equation}
\chi=L^D {\langle \vec{\cal M }^2 \rangle_e}=\hat C(0)\,.\label{chi-micro-def}
\end{equation}

For the specific case of the Ising model, the traditional definitions, using
$S_i=\pm1$ (recall that $s_i=\pm1/\sqrt{2}$) are related with those of the general model through:
\begin{eqnarray}
{\cal U}^\text{Ising}&=&-\sum_{<i,j>}S_i S_j=2{\cal U}-3L^D\;,\nonumber\\
\beta^\text{Ising}&=&\beta/2\;,\label{conversions}\\
\chi^\text{Ising}&=&2\chi\;.\nonumber
\end{eqnarray}

Notice that in $D=2$ this model undergoes a phase transition in $\beta_c=\log(1+\sqrt{Q})$
which is second order for $Q\le4$ and first order for $Q>4$~\cite{Wu}.

\section{Simulation Details}
\label{SIMU-SECT}
We have simulated systems of several sizes in a suitable range of energies
(see Table~\ref{SIMU}).
To update the spins we used a Swendsen-Wang (SW) version of the microcanonical cluster
method~\cite{VICTORMICRO}. This algorithm depends on a tunable
parameter, $\kappa$, which should be as close as possible to
$\langle \hat\beta \rangle_e$ in order to maximize the acceptance of the
SW attempt (SWA). This requires a start-up using a much slower Metropolis
algorithm for determining $\kappa$. In practice, we performed cycles 
consisting of $2\times10^3$ Metropolis steps, $\kappa$ refreshing,
$2\times10^3$ SWA, and a new $\kappa$ refreshing.  We require an acceptance
exceeding $60\%$ to finish these pre-thermalization cycles fixing $\kappa$ for
the following main simulation where only the cluster method is used.

In both studied cases, we have observed a very small autocorrelation time for
all energy values at every lattice size. In the largest lattice for the four
states Potts model we have also consider different starting configurations:
hot, cold and mixed (strips). Although the autocorrelation time is much
smaller, for safety we decided to discard the
first 10\% of the Monte Carlo history using the last 90\% for taking
measurements.

\begin{table}[!ht]
\centering
\begin{tabular*}{\columnwidth}{@{\extracolsep{\fill}}crcrc}
\hline\hline
Model 
&\multicolumn{1}{r}{$L$} 
&\multicolumn{1}{r}{$N_\text{m}(\times 10^6)$}
&\multicolumn{1}{c}{$N_\text{e}$} 
&\multicolumn{1}{c}{Energy range}\\
\hline
$Q=2$, $D=3 $&8    & 20 & 42 &  $[-0.8, -0.9]$ \\
&12   & 20 & 42 &  $[-0.8, -0.9]$ \\
&16   & 20 & 49 &  $[-0.8, -0.9]$ \\
&24   & 20 & 25 &  $[-0.845, -0.875]$ \\
&32   & 20 & 16 &  $[-0.87, -0.860625]$ \\
&48   & 20 & 10 &  $[-0.87, -0.860625]$ \\
&64   & 5  & 10 &  $[-0.870625, -0.865]$ \\
&96   & 5  & 10 &  $[-0.870625, -0.865]$ \\
&128  & 5  & 7  &  $[-0.869375, -0.865625]$ \\
\hline
$Q=4$, $D=2$ &32   & 1024 & 61 &  $[-1.2, -0.9]$ \\
&64   & 128  & 61 &  $[-1.2, -0.9]$ \\
&128  & 32   & 41 &  $[-1.08, -0.98]$ \\
&256  & 32   & 24 &  $[-1.08, -1.005]$ \\
&512  & 25.6 & 32 &  $[-1.07, -1.01]$ \\
&1024 & 6.4  & 30 &  $[-1.06, -1.02]$ \\
\hline\hline
\end{tabular*}
\caption{Simulation details for the two considered models. For each lattice
  size $L$ we show the number of measurements $N_\text{m}$ at each energy
  and the total number of simulated energies uniformly distributed in the
  displayed energy range $N_\text{e}$. For the $Q\!=\!4$, $D\!=\!2$ model, the value of
  $N_\text{m}$ reported have been reached only at specific energies
  near the peaks of the Maxwell construction. Also, additional non-uniformly
  distributed energy values have been simulated near the peaks.
}
\label{SIMU}
\end{table}

\section{Results for the $\bm{D=3}$ Ising model}
\label{ISING3D-SECT}

\begin{figure}[!ht]
\includegraphics[height=\columnwidth,angle=270,trim=43 50 25 25]{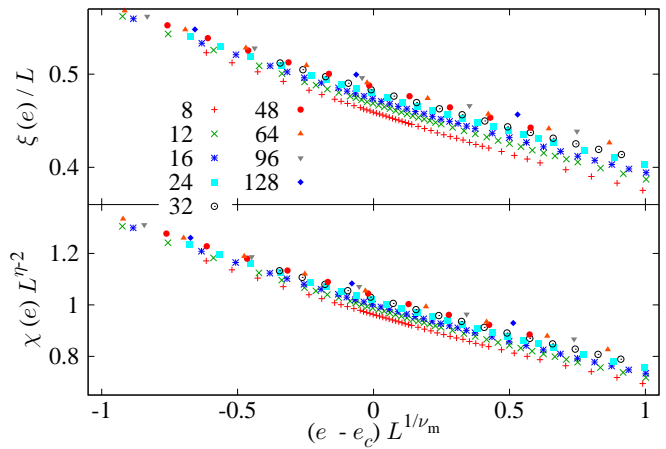}
\caption{(color online). Scaling plot of the correlation length (in
  lattice size units) and the scaled susceptibility for the
  three-dimensional Ising model. We used the critical values,
  $e_\text{c}=-0.867433$ and $\nu_\text{m}=0.7077$. Notice the strong
  scaling corrections for the small systems, as well as the data
  collapse for the largest lattices.  }
\label{scaling_ising}
\end{figure}

In Fig.~\ref{scaling_ising} (upper panel) we show a scaling plot of the correlation length
(in lattice size units) against $(e-e_\text{c}) L^{1/\nu_\text{m}}$. For the
susceptibility we plot $\chi\sim L^{2-\eta}$ (lower panel). If data
followed the expected asymptotic critical behavior with microcanonical
critical exponents they should collapse in a single curve.
In Fig.~ \ref{scaling_ising} we have used the canonical critical quantities
from Refs.~\onlinecite{HASEN, Ising3D} transformed to the microcanonical
counterparts using Eq.~\eqref{EXPO-MICRO-CANONICO}.
From the plot it is clear that important scaling corrections exist in both
cases for the smallest lattices although they are mainly eliminated in the biggest systems.

\begin{figure}[ht!]
\includegraphics[height=\columnwidth,angle=270,trim=28 75 14 25]{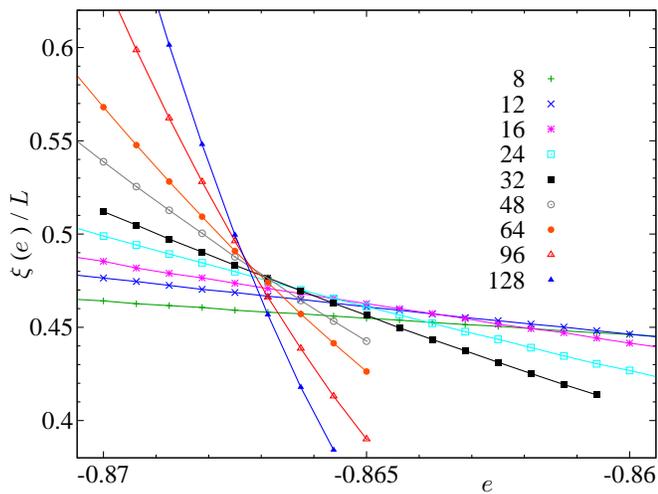}
\caption{(color online). Crossing points of the correlation length in lattice size units
for the three dimensional Ising model. The error bars are in every case smaller than the point sizes.
The values of the different quantities at the crossing as well as the critical exponents are
shown in Table~\ref{table_exp_ising}.} 
\label{fig_xi_ising}
\end{figure}

To obtain the microcanonical critical exponents we used the quotients method,
see Sect.~\ref{subsec:quotients}.
The clear crossing points of the correlation length for different lattice sizes can be seen
in Fig.~\ref{fig_xi_ising}.  The determination of the different quantities at
the crossings, and the position of the crossing itself, requires to interpolate the data 
between consecutive simulated energies (see Fig.~\ref{fig_xi_ising}).
We have found that the method of choice, given de high number of energy values
available, is to fit, using the least squares method,  a selected number of points near the crossing to a
polynomial of appropriate degree. Straight lines do not provide good enough
fits, however, second and third order polynomials give compatible results. In
practice, we have fitted a second order polynomial using the nine nearest
points to the crossing, also comparing the results with those using the seven
nearest points that turn out fully compatible. For error determination we have
always used a jackknife procedure.

\begin{table}[!ht]
\centering
\begin{tabular*}{\columnwidth}{@{\extracolsep{\fill}}rllll}
\hline\hline
$L$ 
& \multicolumn{1}{c}{$e_{\text{c},L,2L}$}
& \multicolumn{1}{c}{$\xi_{L,e_{\text{c},L,2L}}/L$} 
& \multicolumn{1}{c}{$\nu_\text{m}$} 
& \multicolumn{1}{c}{$\eta_\text{m}$} \\
\hline

8   & $-0.861831(12)$ &    0.44922(3)    &   0.8033(42)  &  0.0564(2)  \\
12  & $-0.865010(10)$ &    0.46106(5)    &   0.7968(31)  &  0.0492(4)  \\
16  & $-0.866020(6)$  &    0.46710(5)    &   0.7717(22)  &  0.0469(4)  \\
24  & $-0.866767(3)$  &    0.47411(4)    &   0.7665(11)  &  0.0437(3)  \\
32  & $-0.867034(4)$  &    0.47813(6)    &   0.7594(13)  &  0.0425(5)  \\
48  & $-0.867228(2)$  &    0.48278(5)    &   0.7492(5)   &  0.0412(3)  \\
64  & $-0.867302(2)$  &    0.48555(11)   &   0.7457(16)  &  0.0397(8)  \\\cline{1-5}
\hline\hline
\end{tabular*}
\caption{Lattice size dependent estimates of critical quantities for
  the microcanonical $D=3$ Ising model. The displayed quantities are:
  crossing points $e_{\text{c},L,2L}$ for the correlation length in units of the lattice
  size, $\xi/L$ itself at those crossing points, and the estimates for the
  correlation length exponent $\nu_\text{m}$ and the anomalous
  dimension $\eta$. All quantities are obtained
  using parabolic interpolations.}
\label{table_exp_ising}
\end{table}

The numerical estimates for $e_\text{c}$, $\xi_{L,e_\text{c}}/L$ and
the critical exponents $\nu_\text{m}$ and $\eta$, obtained using the
quotients method for pair of lattices $(L,2L)$ are quoted in
Table~\ref{table_exp_ising}. Our small statistical errors allow to
detect a tiny $L$ evolution. An extrapolation to infinite volume is
clearly needed.

Before going on, let us recall our expectations as obtained applying Fisher
renormalization to the most accurate determination of {\em canonical}
critical exponents known to us [$\nu_\text{m}=\nu/(1-\alpha)=\nu/(D\nu-1)$]:
\begin{eqnarray}
\nu_\text{m}&=& 0.7077(5)\ \text{(from $\nu=0.6301(4)$~\cite{PELI-REP})}\,,\label{NU-FETEN}\\
\eta_\text{m}&=&\eta=0.03639(15)\text{~\cite{CAMPOSTRINI}}\,,\label{ETA-FETEN}\\
\omega&=&0.84(4)\text{~\cite{PELI-REP}}\,.\label{OMEGA-FETEN}
\end{eqnarray}
Besides, although non-universal, let us quote $e_\text{c}=-0.867433(12)$
\footnote{For the $3D$ Ising model at criticality,
  $u_\text{c}^\text{Ising}\!=\!-0.990627(24)$~\cite{HASEN},
  and $\beta_\text{c}^\text{Ising}\!=\!0.2216546(2)$~\cite{Ising3D},
  we obtain for our Potts representation of the Ising model
  $e_\text{c}=(u_\text{c}^\text{Ising}-D)/2+1/(4\beta_\text{c}^\text{Ising})$.
}.

The results obtained from a extrapolation using only leading order
scaling corrections were:
\begin{itemize}
\item $e_\text{c}=-0.867397(6)$, $\omega+1/\nu_\text{m}=1.918(26)$\\ (we
obtained a good fit for $L\geq L_\text{min}=12$, 
with $\chi^2/\text{dof}=0.39/3$,
C.L.=94\%, where ``dof'' stands for \emph{degrees of freedom} and ``C.L.'' for
{\em confidence level}~\footnote{The confidence level is the probability that 
$\chi^2$ would be bigger than the
observed value, supposing that the statistical model is correct. As
a rule, we consider a fit not good-enough whenever C.L.$<10$\%.}).
\item $\xi_{e_\text{c},L}/L=0.5003(12)$, $\omega=0.581(27)$\\ ($L_\text{min}=12$, $\chi^2/\text{dof}=0.12/3$, C.L.=99\%).
\item $\nu_\text{m}=0.714(28)$, $\omega=0.53(30)$\\ ($L_\text{min}=8$, $\chi^2/\text{dof}=3.16/4$, C.L.=53\%).
\item $\eta=0.0391(15), \omega=1.21(24)$\\ ($L_\text{min}=8$, $\chi^2/\text{dof}=0.96/4$, C.L.=92\%).
\end{itemize}
The main conclusions that we draw from these fits are: (i) the
exponents are compatible with our expectations from
Fisher-renormalization, (ii) sub-leading scaling corrections are
important given the tendency of the fits to produce a too low estimate
for $\omega$ (see below) and (iii) the estimates from canonical
exponents (obtained themselves by applying the high-temperature expansion
to improved Hamiltonians~\cite{CAMPOSTRINI,PELI-REP}) are more accurate
than our direct computation in the Microcanonical ensemble.

We can, instead, take an opposite point of view.  If we take the
central values in
Eqs.~(\ref{NU-FETEN},\ref{ETA-FETEN},\ref{OMEGA-FETEN}), as if they
were exact, we can obtain quite detailed information on the amplitudes
for scaling corrections:
\begin{itemize}
\item We find an excellent fit to $\nu_\text{m}(L,2L)=\nu_\text{m} +
  A_1 L^{-\omega} + A_2 L^{-2\omega}$, for
  $L_\text{min}=16$: $\chi^2/\text{dof}=1.53/3$, C.L.=68\%,
  with $A_1=1.38(7)$ and $A_2=-7.6(1.1)$. This
  confirms our suspected strong subleading corrections. Indeed,
  according to these amplitudes $A_1$ and $A_2$, only for $L\approx
  130$ the contribution of the (sub-leading) quadratic term becomes a
  $10\%$ of that of the leading one.
\item In the case of $\eta(L,2L)=\eta + B_1 L^{-\omega}+ B_2
  L^{-2\omega}\,$, for $L_\text{min}=8$: $\chi^2/\text{dof}=2.4/5$, C.L.=79\%,
  we have $B_1=0.101(10)$ and $B_2=0.07(7)$. Subleading scaling corrections
  are so small that, within our errors, it is not clear whether $B_2=0$ or not.
\end{itemize}

The quite strong scaling corrections found for $\nu_\text{m}$ may cast
some doubts in the extrapolation for $\xi_{L,e_\text{c}}/L$, the only
quantity that we cannot double-check with a canonical computation. To control
this, we proceed to a fit including terms linear and quadratic in
$L^{-\omega}$ with $\omega=0.84(4)$. We get
$$\frac{\xi_{L,e_\text{c}}}{L}=0.4952(5)(7),$$
with $L_\text{min}=12$, $\chi^2/3=2.17/3$, C.L.=54\%.
Here, the second error is due to the quite small uncertainty in
$\omega$. It is remarkable that the contribution to the error stemming
from the error in $\omega$ is {\em larger} than the purely statistical
one.

\section{Results for the $\bm{D=2}$, $\bm{Q=4}$ Potts Model}
\label{POTTS2D-SECT}

The $Q=4$ $D=2$ Potts model offers two peculiarities that will be
explored here.  First, it suffers from quite strong logarithmic
scaling corrections. And second, it displays pseudo
metastability~\cite{Fukugita}, an ideal playground for a microcanonical
study.

The study of the FSS for the $Q=4$, $D=2$ Potts
model~\cite{Salas}, based on the analysis of the Renormalization Group (RG)
equations~\cite{Cardy}, reveals the presence of multiplicative scaling
corrections. This is one of the possible forms that scaling
corrections can take in the limit $\omega\to 0$, and is a great
nuisance for numerical studies. A very detailed theoretical input is
mandatory to perform safely the data analysis. We shall make here an
educated guess for the {\em microcanonical} form of the scaling corrections,
based purely in ensemble-equivalence and in the {\em canonical} results.

From ensemble-equivalence we expect
\begin{equation}	
e -e_\text{c} \sim C(L,\beta_\text{c}) \Delta \beta_L \ ,
\end{equation}
where $C(L,\beta_\text{c})$ is the finite-lattice canonical specific
heat at $\beta_\text{c}$ and $\Delta \beta=
\beta_\text{c}^{(L)}-\beta_\text{c}$ is the inverse-temperature
distance to the critical point of any $L$-dependent feature (such as
the temperature maximum of the specific-heat, etc.). We borrow from
Ref.~\cite{Salas} the leading FSS behavior for these quantities:
\begin{equation}	
C(L,\beta_\text{c}) \sim \frac{L}{(\log L)^{3/2}} \quad , \quad \Delta \beta_L \sim \frac{(\log L)^{3/4}}{L^{3/2}}\,. 
\end{equation}
Thus, we have:
\begin{equation}	
e(L) -e_\text{c}(\infty) \sim L^{-1/2} (\log L)^{-3/4} \ .
\label{potts_energy_form}
\end{equation}
This result can be derived as well by considering only the leading
terms of the first derivative of the singular part of free energy
respect to the thermal field, $\phi(\propto
\beta-\beta_\mathrm{c})$~\cite{Salas}:
\begin{align}
\frac{\partial f_\text{sing}(\phi,h,\psi)}{\partial \phi} \approx \frac{4}{3}D_\pm |\phi|^{1/3} (-\log |\phi|)^{-1}+\notag \\
D_\pm |\phi|^{4/3} (-\log |\phi|)^{-2} \frac{1}{\phi} \ ,
\end{align}
The previous equation describes the energy of the
system and its leading term is
\begin{equation}	
e - e_\text{c} \sim \frac{4}{3}D_\pm \frac{|\phi|^{1/3}}{\log |\phi|} \ ,
\end{equation}
but 
\begin{equation}	
\phi \approx C'_\pm L^{-3/2} (\log L)^{3/4} \ ,
\end{equation}
so it is direct to obtain again Eq.~\eqref{potts_energy_form}.  Hence, 
we are compelled to rephrase Eq.~\eqref{FSSA2} as
\begin{equation}
\langle O\rangle_{L,e}= L^{\frac{x_{O,\text{m}}}{\nu_\text{m}}} \tilde f_O\left (L^{1/2} (\log L)^{3/4} (e-e_\text{c})\right)+\ldots\,.\label{FSSA-POTTS}
\end{equation}
Furthermore, from the canonical analysis~\cite{Salas}, we expect
multiplicative logarithmic corrections to the susceptibility (that
do not get Fisher renormalized). %VMM: creo que es asi. True?
Furthermore, the dots in~\eqref{FSSA-POTTS} stand for 
corrections of order $\log \log L / \log L $ and $1/\log L$~\cite{Salas}.

We first address in Sect.~\ref{subsec:exponents} the direct
verification of Eq.~\eqref{FSSA-POTTS} using the quotients method. We
consider afterwards the pseudo-metaestability features.

\subsection{Scaling Plots and Critical Exponents}
\label{subsec:exponents}

\begin{figure}[!ht]
\includegraphics[height=\columnwidth,angle=270,trim=68 58 26 25]{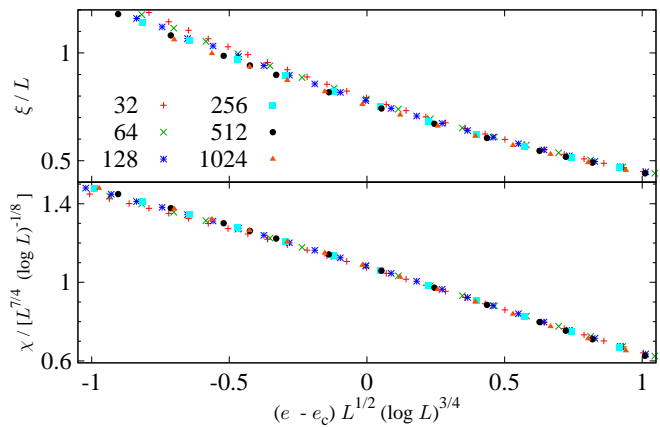}
\caption{(color online). Graphical demonstration of
  Eq.~\eqref{FSSA-POTTS} as applied to the microcanonical $D=2$, $Q=4$
  Potts model: both the correlation length in units of the lattice
  size ({\bf top}) and the scaled susceptibility, $\overline{\chi}$ in
  Eq.~\eqref{inv_chi} ({\bf bottom}), are functions of the scaling
  variable $(e-e_\text{c})L^{1/2}(\log L)^{3/4}$.}
\label{scaling_xiandsus_pottsmicro}
\end{figure}

We start by a graphical demonstration of Eq.(\ref{FSSA-POTTS}):
$\xi/L$ as a function of $(e-e_c)L^{1/2}(\log L)^{3/4}$, should
collapse onto a single curve (the deviation will be bigger for small
$L$ values, due to neglected scaling corrections of order $\log \log L
/ \log L $ and $1/\log L$)~\footnote{
We obtain the exact $e_\text{c}$ in the thermodynamic limit from
$\beta_\text{c}=\log (1+\sqrt{Q})$~\cite{Baxter}, 
and  $u_\text{c}=-(1+Q^{-1/2})$~\cite{Wu} by applying
$e_\text{c}=u_\text{c}+1/(2\beta_\text{c})$}.  A similar behavior is expected
for the scaled susceptibility~\cite{Salas}:
\begin{equation}
\overline{\chi}=\frac{\chi}{L^{7/4}(\log L)^{-1/8}} \,.
\label{inv_chi}
\end{equation}
Note that $\xi/L$ does not need an additional logarithmic factor.
These expectations are confirmed in
Fig.~\ref{scaling_xiandsus_pottsmicro}, specially for the largest
system sizes (that suffer lesser scaling corrections).

\begin{figure}[!ht]
\includegraphics[height=\columnwidth,angle=270,trim=13 58 6 25]{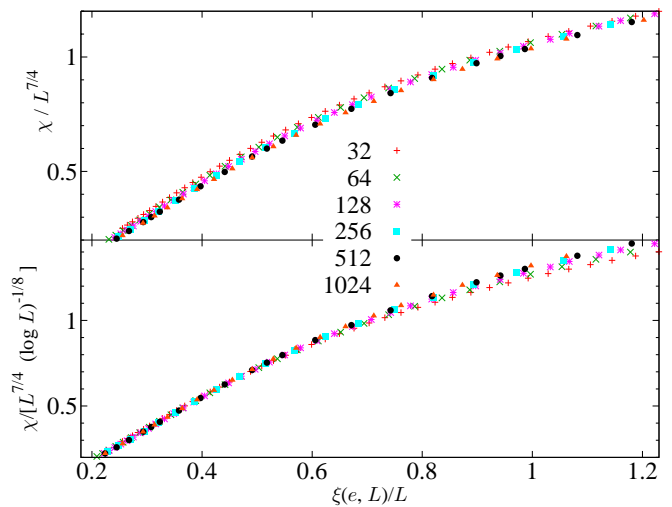}
\caption{(color online). Comparison of the scaling for the naively
  scaled susceptibility $\chi L^{-7/4}$ ({\bf top}) and for
  $\overline{\chi}$ ({\bf bottom}), as a function of the correlation
  length in units of the lattice size, for the microcanonical $D=2$,
  $Q=4$ Potts model.  }
\label{scaling_susVSxi_pottsmicro}
\end{figure}

We can check directly the importance of the multiplicative logarithmic
corrections for the susceptibility by comparing $\chi$ and $\overline{\chi}$
as a function of $\xi/L$, Fig.~\ref{scaling_susVSxi_pottsmicro}. The improved scaling of $\overline{\chi}$ is apparent. We observe as well  
that the largest corrections to scaling are found at and
below the critical point (around $\xi/L \approx 1.0$).

We now proceed to the numerical computation of critical exponents. We
shall use the quotients method, modified as described in
Appendix~\ref{AppendixA}.  As it is evident from
Fig.~\ref{fig_cortes}, the crossing points can be obtained with great
accuracy using parabolic interpolations of the nine points around the
estimated crossing energies, see Sect.~\ref{MODELS_OBS-SECT}.  We
checked that the results do not depend on the interpolating polynomial
degree by comparing with interpolations using cubic curves. We also
compared with the results obtained using only seven points around the
crossing obtaining again full agreement.

The obtained critical exponents are shown in Table~\ref{table_exp}, we
may compare them with the exact ones~\cite{Wu} ($\nu=2/3$, $\alpha=2/3$ and
$\eta=1/4$):
\begin{equation}
\nu_\text{m}=2 \quad \quad ;\quad \quad \eta=\eta_\text{m}=\frac{1}{4}\, .
\end{equation}
Comparing with our computed exponents we
obtain an acceptable agreement.  In the case of the microcanonical
$\nu$ exponent, $\nu_\text{m}$, after adding the correction for the
quotients method in presence of logarithms, the agreement is fairly good. We can see a clear
trend towards the exact result value for all the lattice sizes except
for the biggest one (2.5 standard deviations away), which is probably
due to a bad estimation of the huge temperature derivatives of the
correlation length.  In the case of the microcanonical $\eta$
exponent, $\eta_\text{m}$, which must be the same that the canonical
one, we can see clearly the tendency to the analytical value
$\eta_\text{m}=0.25$.  We must remark the importance of adding the
corrections described in Appendix~\ref{AppendixA} to the quotients
method.

\begin{figure}[!ht]
\includegraphics[height=\columnwidth,angle=270,trim=28 67 13 25]{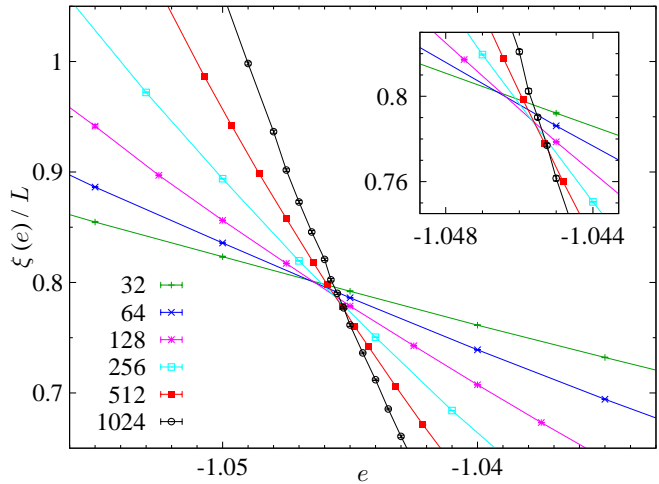}
\caption{(color online). Correlation length in lattice size units for
  the two-dimensional $Q=4$ Potts model.  The values of the different
  quantities on the crossings for lattices $L$ and $2L$, as well as
  the corresponding estimate for critical exponents, are in
  Table~\ref{table_exp}. The inset is a magnification of the critical
  region.  }
\label{fig_cortes}
\end{figure}

\begin{table*}[t]
\centering
\begin{tabular*}{\textwidth}{@{\extracolsep{\fill}}rllllll}
\hline\hline
\multicolumn{1}{c}{$L$} 
&\multicolumn{1}{c}{$e_{\text{c},L,2L}$} 
&\multicolumn{1}{c}{$\xi_{L,e_{\text{c},L,2L}}/L$} 
&\multicolumn{1}{c}{$\nu_\text{m}$} 
&\multicolumn{1}{c}{$\nu_\text{m}'$} 
&\multicolumn{1}{c}{$\eta_\text{m}$} 
&\multicolumn{1}{c}{$\eta_\text{m}'$} \\
\hline
32  & $-1.04659(5)$ &  0.8016(5)   &  1.534(6)  &    1.998(10)  &    0.2663(9)   &   0.2334(9)    \\
64  & $-1.04633(2)$ &  0.7990(3)   &  1.554(8)  &    1.957(12)  &    0.2638(6)   &   0.2360(6)    \\
128 & $-1.04579(1)$ &  0.7909(3)   &  1.578(5)  &    1.938(7)   &    0.2639(5)   &   0.2398(5)    \\
256 & $-1.04548(2)$ &  0.7836(5)   &  1.643(12) &    1.987(17)  &   0.2615(11)   &   0.2402(11)   \\
512 & $-1.04519(2)$ &  0.7734(9)   &  1.602(31) &    1.895(42)  &   0.2617(21)   &   0.2427(21)	\\
\hline\hline
\end{tabular*}
\caption{Crossing points of the correlation length in lattice size units as a function of the energy
for pairs of lattices \hbox{($L$, $2L$)}.  Using the original quotients method~\cite{VICTORAMIT} we obtain the microcanonical critical exponents, shown in the columns 4 and 6, while the corrected ones (columns 5 and 7) are labelled with primed symbols, see Appendix~\ref{AppendixA}.}
\label{table_exp}
\end{table*}

\subsection{Critical point, latent heat and surface tension}
\label{subsec:temperature}

\begin{figure}[!ht]
\includegraphics[height=\columnwidth,angle=270,trim=28 76 14 25]{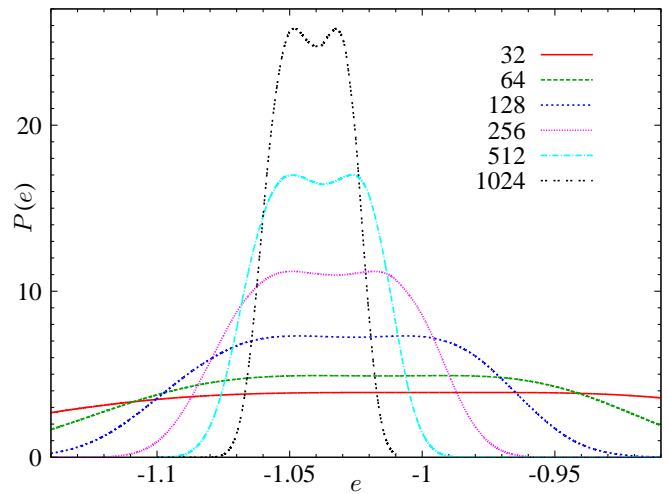}
\caption{(color online).  {\em Canonical} probability distribution
  function for the energy density, $P_{\beta}^{(L)}(e)$, as
  reconstructed from microcanonical simulations of the $D=2$, $Q=4$
  Potts model and different system sizes. The $L$ dependent critical
  point $\beta_{\text{c},L}$ is computed using the Maxwell rule,
  Sect.~\ref{SECT-DOS-PICOS} (note the equal height of the two peaks
  enforced by Maxwell's construction). The system displays an apparent
  latent heat, that becomes smaller for growing $L$, and vanish in the
  large $L$ limit.
} 
\label{HISTOGRAMAS}
\end{figure}

It has been known for quite a long time that the $D=2$, $Q=4$ Potts
model on finite lattices show features typical of first-order phase
transitions~\cite{Fukugita}. For instance, see Fig.~\ref{HISTOGRAMAS},
the probability distribution function for the internal energy,
$P_\beta(e)$, display two peaks at energies $e_\text{d}$ (the
coexisting {\em disordered} phase) and $e_\text{o}$ (the energy of the
{\em ordered} phase) separated by a minimum at $e^*$. Of course, since
the transition is of the second order, $e_\text{c}$ is the common
large $L$ limit of $e_\text{d}$, $e_\text{o}$ and $e^*$.

\begin{figure}[!ht]
\includegraphics[height=\columnwidth,angle=270,trim=23 100 0 25]{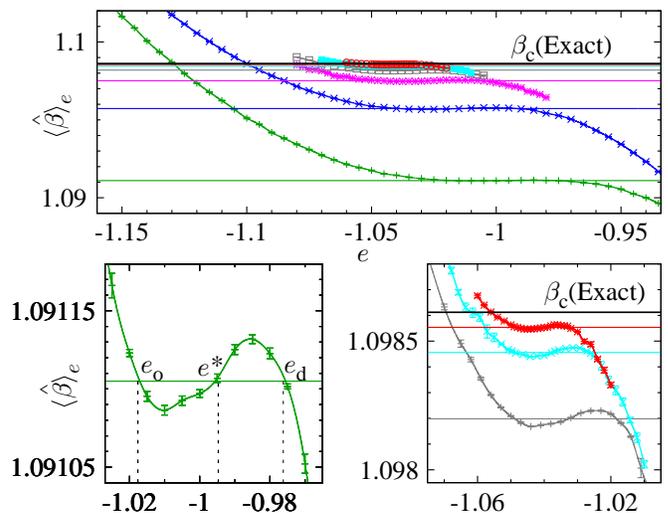}
\caption{(color online).{\bf Top:} From the microcanonical mean values
  $\langle \hat\beta\rangle_{e,L}$ for the $D=2$, $Q=4$ Potts model, we
  estimate the size dependent {\em canonical} inverse critical
  temperature $\beta_{\text{c},L}$ (horizontal lines) for all the
  simulated lattice sizes, ranging from $L=32$ (lower) to $L=1024$
  (upper). We show as well as the analytical prediction (upper
  horizontal line).  {\bf Bottom-left:} example of Maxwell construction
  for our $L=32$ data. The $e$-integral of $\langle
  \hat\beta\rangle_{e,L}-\beta_{\text{c},L}$ from $e_\text{o}$ to
$e_\text{d}$ vanish. {\bf Bottom-right}: zoom of upper panel showing
only data for lattice sizes $L=256$ (lower curve), $L=512$ (medium curve) and
$L=1024$ (upper curve). }
\label{beta_e_L}
\end{figure}

We discussed in Sect.~\ref{SECT-DOS-PICOS} how the Maxwell construction is
used to estimate the canonical critical point $\beta_{\text{c},L}$, as
well as $e_\text{d}$, $e_\text{o}$ and the associated surface tension.
This procedure is outlined in Fig.~\ref{beta_e_L}. The numerical results
are in Table~\ref{numerical_beta}, where we see that 
$\beta_{\text{c},L}$ is a monotonically increasing
function of $L$ continuously approaching to the analytical value
$\beta_\text{c}= \log(1+\sqrt{Q})=1.0986122\ldots$~\cite{Baxter}. A jackknife
method~\cite{VICTORAMIT} is used to compute the error bars for all quantities
in Table~\ref{numerical_beta}.

\begin{table}[!ht]
\centering
\begin{tabular*}{\columnwidth}{@{\extracolsep{\fill}}rllll}
\hline\hline
\multicolumn{1}{c}{$L$} 
& \multicolumn{1}{c}{$\beta_{\text{c},L}$} 
& \multicolumn{1}{c}{$e_\text{o}$} 
& \multicolumn{1}{c}{$e_\text{d}$} 
& \multicolumn{1}{c}{$\varSigma \times 10^5$} \\\hline
32    & 1.0911070(20) &  -1.0175(4) & -0.9760(2)     &  0.47(2) \\
64    & 1.0957256(14) &  -1.0392(3) & -0.9915(2)     &  2.77(7) \\
128   & 1.0975150(10) &  -1.0463(3) & -1.0062(5)     &  4.10(15)\\
256   & 1.0981989(5)  &  -1.0489(2) & -1.0183(3)     &  3.92(8) \\
512   & 1.0984570(3)  &  -1.0490(1) & -1.0266(2)     &  3.28(11)\\
1024  & 1.0985539(3)  &  -1.0483(3) & -1.0325(1)     &  2.09(17)\\
\hline\hline
\end{tabular*}
\caption{Using Maxwell construction, we compute for the $D=2$, $Q=4$ Potts
model the $L$-dependent estimate of the (inverse) critical temperature
$\beta_{\text{c},L}$, the energies of the coexisting ordered phase $e_\text{o}$,
 and disordered phase $e_\text{d}$, as well as the surface tension ($\varSigma$).}
\label{numerical_beta}
\end{table}

To perform a first check of our data, we observe that
$\beta_{\text{c},L}$ is a typical {\em canonical} estimator of the inverse
critical temperature. As such, it is subject to standard canonical FSS,
where the main scaling corrections come from two additive logarithmic
terms~\cite{Salas}:
\begin{align}
\beta_{\text{c},L}-\beta_\text{c}=a_1 \frac{(\log L)^{3/4}}{L^{3/2}} \times \qquad \qquad \qquad \notag \\
\left(1+a_2\frac{\log\log L}{\log L}+a_3 \frac{1}{\log L} \right) \;.
\end{align}
From our data in Table~\ref{numerical_beta}, we obtain
$a_1=-0.44(7)$, $a_2=-1.15(72)$, and $a_3=2.28(26)$, 
and a good fit ($L_\text{min}=128$: $\chi^2/\text{dof}=0.28/1$, C.L.=60\%). 

\begin{figure}[!ht]
\includegraphics[height=\columnwidth,angle=270,trim=28 83 22 25]{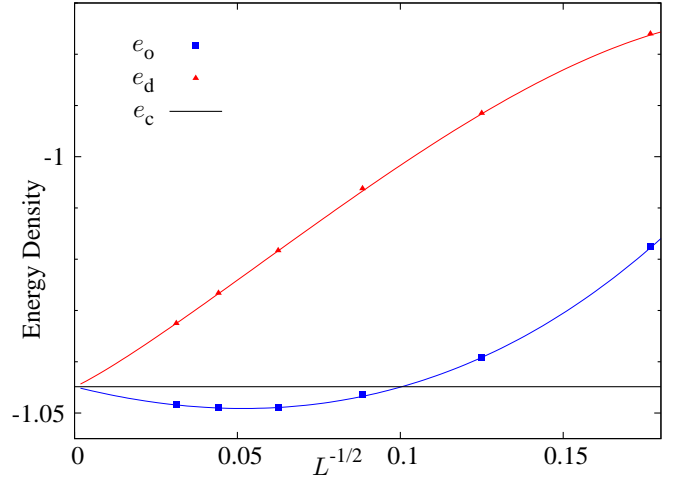}
\caption{(color online). System size dependent estimates for the
  energies of the ``coexisting'' ordered ($e_\text{o}$, blue points)
  and disordered ($e_\text{d}$, red points) phases of the $D=2$, $Q=4$
  Potts model, as a function of $L^{-1/2}$. The lines are fits to the
  expected analytical behavior Eq.~(\ref{scaling_corr}). The
  horizontal line corresponds to the asymptotic value, $e_\text{c}$.  }
\label{extrap_ener_sokal}
\end{figure}

As for the $L$ dependence of $e_\text{d}$ and $e_\text{o}$, we try a
fit that consider the expected scaling correction terms~\cite{Salas}: 
\begin{align}
e_{\text{c,o},L} -e_c = a_1 L^{-1/2} (\log L)^{-3/4} \times \qquad \notag \\
\left(1+  a_2 \frac{\log \log L}{\log L}+{a_3}\frac{1}{\log L}\right) .\label{scaling_corr}
\end{align}
Our results for $e_\text{o}$ are: $a_\text{1o}=-2.03(20)$, $a_\text{2o}=-1.65(27)$,
and $a_\text{3o}=-2.08(41)$, with a fair fit quality
($L_\text{min}\!=\!32$: $\chi^2/\text{dof}=2/3$, C.L.=57\%). On the other hand,
we
obtain for $e_\text{d}$: $a_\text{1d}=2.02(14)$,
$a_\text{2d}=0.93(37)$, and $a_\text{3d}=-2.93(34)$ , with a fair fit
as well
($L_\text{min}\!=\!32$: $\chi^2/\text{dof}=0.84/3$, C.L.=84\%). 
These two fits are shown in  Fig.~\ref{extrap_ener_sokal}.

For the surface tension, we note in Table~\ref{numerical_beta} a non monotonic
behavior. Furthermore, we lack a theoretical input allowing us to fit. We thus
turn to a variant of the quotients method. Where $\varSigma$ to follow a pure
power law scaling, $\varSigma\propto L^b$, exponent $b$ would be obtained as:
\begin{equation}
\frac{\varSigma(L_1)}{\varSigma(L_2)}= \left( \frac{L_1}{L_2} \right) ^b
\Longrightarrow
b= \frac{\log [\varSigma(L_1)/\varSigma(L_2)] }{\log (L_1/L_2)  }
\label{eff_exp}
\end{equation}
The effective exponent $b$ obtained from our data is displayed in Table~\ref{table_eff_exp}. We observe that it is clearly negative (as it should since
$\varSigma$ vanishes for a second order phase transition). An asymptotic
estimate, however, seems to require the simulation of larger systems.
\begin{table}[!ht]
\centering
\begin{tabular*}{\columnwidth}{@{\extracolsep{\fill}}cl}
\hline\hline
$(L_1,L_2)$ 
& \multicolumn{1}{c}{$b_\text{eff}(\varSigma)$}\\
\hline
(32,64)    &  $\hphantom{-}2.56(7)$ \\
(64,128)   &  $\hphantom{-1}0.56(6)$ \\
(128,256)  &  $-0.065(60)$ \\
(256,512)  &  $-0.257(57)$ \\
(512,1024) &  $-0.650(127)$ \\
\hline\hline
\end{tabular*}
\caption{Effective exponent obtained using Eq.~\eqref{eff_exp} for the  surface tension.}
\label{table_eff_exp}
\end{table}

\begin{table*}[!ht]
\centering
\begin{tabular*}{\textwidth}{@{\extracolsep{\fill}}rllllll}
\hline\hline
\multicolumn{1}{c}{$L$} 
& \multicolumn{1}{c}{$\xi(e_\text{o})/L$} 
& \multicolumn{1}{c}{$\xi(e_\text{d})/L$} 
& \multicolumn{1}{c}{\raisebox{2pt}{$\overline{\chi}(e_\text{o})$}} 
& \multicolumn{1}{c}{\raisebox{2pt}{$\overline{\chi}(e_\text{d})$}}
& \multicolumn{1}{c}{$\xi^\text{canonical}/L$} 
& \multicolumn{1}{c}{\raisebox{2pt}{$\overline{\chi}^\text{canonical}$}} \\\hline
32    & 0.637(2)  & 0.453(1)  & 0.990(3) & 0.907(2)  & 0.647(1)  & 1.287(3)  \\
64    & 0.732(3)  & 0.396(1)  & 0.995(2) & 1.025(3)  & 0.545(2)  & 1.310(2)  \\
128   & 0.799(5)  & 0.357(4)  & 1.001(3) & 1.106(5)  & 0.472(7)  & 1.331(3)  \\
256   & 0.866(6)  & 0.335(3)  & 1.001(5) & 1.182(6)  & 0.429(5)  & 1.343(5)  \\
512   & 0.915(4)  & 0.315(2)  & 1.014(8) & 1.238(4)  & 0.392(4)  & 1.366(8)  \\
1024  & 0.953(15) & 0.302(2)  & 0.997(21)& 1.279(13) & 0.367(3)  & 1.353(22) \\
\hline\hline
\end{tabular*}
\caption{Correlation length in units of the lattice size and the
RG invariant   {$\overline{\chi}$} defined in Eq.~(\ref{inv_chi}), for several $L$
values, as computed in the microcanonical $D=2$, $Q=4$ Potts model. The
chosen values of the energy density  correspond to the ordered ($e_\text{o}$) and disordered ($e_\text{d}$) phases. For comparison we also display the {\em canonical} results at $\beta_\text{c}$ obtained in Ref.~\cite{Salas}.}
\label{numerical_corrlenght}
\end{table*}

We have just seen that, up to scaling corrections, $e_\text{d}^{(L)}$
and $e_\text{o}^{(L)}$ correspond to (different) $L$-independent
values of the argument of the scaling function $\tilde f_\xi$ in
Eq.~\eqref{FSSA-POTTS}. Hence we expect that $\xi(e_\text{d})/L$ and
$\xi(e_\text{o})/L$, see table~\ref{numerical_corrlenght}, approach
non-vanishing, different values in the large $L$ limit. The
finite-size scaling corrections are expected to be additive
logarithms~\cite{Salas}
\begin{equation}
\frac{\xi}{L} = a+\frac{b}{\log L}
\end{equation}
The results are
\begin{equation}
\frac{\xi(e_\text{o})}{L}=1.28(1)-\frac{2.28(5)}{\log L}\,,
\end{equation}
($L_\text{min}=32$: $\chi^2/\text{dof}=4.2/3$, C.L.=22\%), and
\begin{equation}
\frac{\xi(e_\text{d})}{L}=0.159(4)-\frac{0.98(2)}{\log L}\,
\end{equation}
($L_\text{min}=32$: $\chi^2/\text{dof}=3.3/3$, C.L.=37\%).

A very similar analysis can be performed for the scaled
susceptibility, Eq.~(\ref{chi-micro-def}), at $e_\text{d}$ and
$e_\text{o}$.  In order to deal with the multiplicative logarithms of
the susceptibility we rather used $\overline{\chi}$ defined in
Eq.~\eqref{inv_chi}.

Fitting our data set to the logarithmic form
\begin{equation}
\overline{\chi}=A+B \frac{\log \log L}{\log L}
\end{equation}
 obtained in Ref.~\cite{Salas}, we obtain a good fit in the ordered phase energy, $e_\text{o}$:
\begin{equation}
\overline{\chi}(e_\text{o})=2.41(5)-4.00(15) \frac{\log \log L}{\log L}\,,
\end{equation}
($L_\text{min}= 128$: $\chi^2/\text{dof}=3.10/2$, C.L.=21\%). On the
other hand the extrapolation for the susceptibility defined in the
disordered phase energy, $e_\text{d}$, is a nonsensical negative
value.

We can also fit the data to the logarithmic form also used in Ref.~\cite{Salas}:
\begin{equation}
\overline{\chi}=A+\frac{B}{\log L}
\end{equation}
finding:
\begin{equation}
\overline{\chi}(e_\text{o})=1.643(5)-\frac{2.55(2)}{\log L}\,,
\end{equation}
($L_\text{min}=32$: $\chi^2=7.44/4$, C.L.=11\%),  and
\begin{equation}
\overline{\chi}(e_\text{d})=0.094(7)+\frac{1.87(37)}{\log L}\,,
\end{equation}
($L_\text{min}=64$: $\chi^2/\text{dof}=2.94/3$, C.L.=37\%)\,.
For comparison, we recall that
Ref.~\cite{Salas} reports two different fits for $\overline{\chi}$, depending
of the logarithmic corrections they used:
\begin{eqnarray}
\overline{\chi}^{\text {canonical}}&=&1.673(33) -1.056(98) \frac{\log \log
    L}{\log L}\,,\\
\overline{\chi}^{\text{canonical}}&=&1.454(13) -\frac{0.600(55)}{\log L} \,.
\end{eqnarray}

\section{Conclusions}
\label{CONC-SECT}
\setcounter{equation}{0}

We have formulated the Finite Size Scaling Ansatz for microcanonical
systems in terms of quantities accessible in a finite lattice. This
form allows to extend the Phenomenological Renormalization approach
(the so called quotients method) to the microcanonical framework.

Our Ansatz has been subjected to a strong numerical testing. We have
performed extensive microcanonical numerical simulations in two
archetypical systems in Statistical Mechanics: the three dimensional
Ising model and the two-dimensional four states Potts model. 
The two models present a power-law singularity in their
canonical specific heat, implying  non-trivial Fisher
renormalization when going to the microcanonical ensemble.
A microcanonical cluster method works for both models, hence allowing us
study very large system sizes ($L=128$ in $D=3$ and $L=1024$ in
$D=2$). 

In the case of the Ising model, we have obtained precise
determinations of the critical exponents, that, we feel, provide
strong evidence for our extended microcanonical FSS Ansatz.

For the Potts model, very strong logarithmic corrections (both
multiplicative and additive) plague our data. Fortunately, we have a
relatively strong command on these corrections from canonical
studies~\cite{Salas}. Our data can be fully rationalized using the scaling 
corrections suggested by the theoretical analysis.

\section*{Acknowledgements}

We have been partly supported through Research Contracts
No. FIS2006-08533-C03 and No. FIS2007-60977 (MICINN, Spain). The
simulations for this work were performed at BIFI.

\appendix

\section{The Quotients Method in presence of multiplicative logarithmic corrections}
\label{AppendixA}

The quotients method~\cite{QUOTIENTS,VICTORAMIT}, as been widely used 
in the past for the computation of critical exponents. Yet,
its convergence to the large $L$ limit is extremely slow in presence of
multiplicative logarithmic scaling corrections. Fortunately,
let us show how we can speed up convergence if we have enough
analytical information at our disposal.

Let us consider an observable $O$ such that its FSS behavior is given by ($z$ can
be either the reduced temperature $t$ or $e-e_\text{c}$)
\begin{equation}
O(L,z)=L^{x_O/\nu}(\log L)^{\widehat{x}_O}\left[ F_O\left(\frac{L}{\xi(L,z)}\right) + \ldots \right] \ ,
\end{equation}
then the critical exponent calculated using Eq.~(\ref{QUO2}) must be corrected following:
\begin{equation}
\frac{x_O'}{\nu}=\frac{x_O}{\nu}-\frac{\widehat{x}_O}
{\log(L_2/L_1)}\log \left( \frac{\log L_2}{\log L_1} \right) \ .
\end{equation}

Specifically for the two dimensional four states Potts model the values of the logarithmic correction exponents are analytically known~\cite{Cardy,Salas} thus we can 
calculate accurately the corrections in this case. In addition, the susceptibility behaves as
\begin{equation}
\chi \sim L^{7/4}(\log L)^{-1/8} 
\label{scaling_chi}
\end{equation}
so we easily get
\begin{equation}
\eta'=\eta-\frac{1}{8 \log(L_2/L_1)}\log \left( \frac{\log L_2}{\log L_1} \right) \ .
\end{equation}
For the correlation length it is known that
\begin{equation}
\xi \sim |t|^{-2/3}(- \log t)^{1/2}  \quad\quad ; \quad\quad  t \sim L^{-3/2}(\log L)^{3/4}
\end{equation}
and therefore his temperature derivative scales as
\begin{equation}
\partial_\beta \xi \sim L^{5/2}(\log L)^{-3/4}
\end{equation}
resulting in a $\nu$ canonical exponent correction of
\begin{equation}
\nu'=\nu\left[1-\frac{3}{4}\frac{\nu}{\log(L_2/L_1)}\log \left( \frac{\log L_2}{\log L_1} \right) \right] \ .
\end{equation}

While for the microcanonical $\nu$ exponent, $\nu_\text{m}$, we use that
\begin{equation}
e \sim L^{-1/2}(\log L)^{-3/4} 
\end{equation}
and
\begin{equation}
\partial_e \xi \sim L^{3/2}(\log L)^{3/4} \;.
\end{equation}
Hence, 
\begin{equation}
\nu'_{_\text{m}}=\nu_{_\text{m}}\left[1+\frac{3}{4}\frac{\nu_{_\text{m}}}{\log(L_2/L_1)}\log \left( \frac{\log L_2}{\log L_1} \right) \right] \ .
\end{equation}

\end{document}